\documentclass[aps,prb,reprint,superscriptaddress,showpacs]{revtex4-2}

\usepackage[utf8]{inputenc}
\usepackage[english]{babel}
\usepackage{amssymb,amsmath,mathrsfs,bm,graphicx,color}
\usepackage[pdftex,colorlinks=true]{hyperref}

\makeatletter\AtBeginDocument{\let\@elt\relax}\makeatother 

\begin{document}

\author{Pascal Thibaudeau}
\email{pascal.thibaudeau@cea.fr}
\affiliation{CEA, DAM, Le Ripault, BP 16, F-37260, Monts, FRANCE}

\author{Stam Nicolis}
\affiliation{Institut Denis Poisson, Université de Tours, Université d'Orléans, CNRS (UMR7013), Parc de Grandmont, F-37200, Tours, FRANCE}
\email{stam.nicolis@lmpt.univ-tours.fr}

\title{Emerging magnetic nutation}
\date{\today}

\begin{abstract}
    Nutation has been recognized as of great significance for spintronics; but justifying its presence has proven to be a hard problem. 
    In this paper we show that nutation can be understood as emerging from a systematic expansion of a kernel that describes the history of the interaction of a magnetic moment with a bath of colored noise. 
    The parameter of the expansion is the ratio of the colored noise timescale to the precession period. 
    In the process we obtain the Gilbert damping from the same expansion. 
    We recover the known results, when the coefficients of the two terms are proportional to one another, in the white noise limit; and show how colored noise leads to situations where this simple relation breaks down, but what replaces it can be understood by the appropriate generalization of the fluctuation--dissipation theorem.
    Numerical simulations of the stochastic equations support the analytic approach.
    In particular we find that the equilibration time is about an order of magnitude longer than the timescale set by the colored noise for a wide range of values of the latter and we can identify the presence of nutation in the non-uniform way the magnetization approaches equilibrium.     
\end{abstract}

\maketitle
\section{Introduction}
\label{intro}
Recent progress in spintronics has led to the search for processes and materials that can realize ever shorter switching times for the magnetization--and this has opened a window to a régime, where nutation effects of the average magnetization cannot be ignored.
How to take them into account becomes, therefore, of practical interest~\cite{neerajInertialSpinDynamics2020}. 
However how to describe the emergence and the relevance of nutation from first principles, as magnetic moments interact with a bath, has been and remains a challenging problem. One reason is that it is by no means obvious how to extract its properties from the interaction with the bath.

For magnetic materials a common way of describing the effects of the bath is by the  so--called Gilbert damping mechanism~\cite{gilbertAnomalousRotationalDamping1955}.
What the two effects have in common is the vector nature of the bath; where they differ is in how this gets imprinted on the magnetization profile in each case. 

Providing a microscopic picture of how Gilbert damping may appear has long been recognized as an outstanding question and there have been many attempts for explaining how it may occur.
However whether there might be any relation with the effects of nutation has only received attention.~\cite{makhfudzNutationWavePlatform2020}, where both were assumed to be present and certain consequences for ultrafast switching were set forth. 
In that case, though, it was assumed that the magnetic moment was not in interaction with a stochastic bath in full generality: the latter was present only indirectly, through the deterministic Gilbert term. 

In this paper we shall show that the nutation term and the Gilbert term can both be obtained as well--defined contributions from a systematic expansion of the equations of motion of a magnetic moment, interacting with a vector bath, whose stochastic component is drawn from colored noise. 
The expansion parameter can be identified as the ratio of the timescale of the colored noise to the precession frequency of the Larmor motion, here simply reduced to a Zeeman field only; this simplifies the calculations, without any loss of generality.

The equation of motion of a classical or quantum magnetic moment, in the presence of this external field, is of first order in the dynamical variables and describes precession~\cite{cohen-tannoudjiQuantumMechanics1993}.
This equation implies, in particular, some non-trivial conservation laws: the norm of the magnetic moment is conserved and, for a constant external field, so is the component along it~\cite{thibaudeauNambuMechanicsStochastic2017,makhfudzNutationWavePlatform2020}. 

When the magnetic moment interacts with a bath the conservation laws take the form of  fluctuation--dissipation relations, that describe the fact that the magnetic moment is in equilibrium with the bath. 
Indeed, the proposal in ref.~\cite{mondalRelativisticTheoryMagnetic2017} describes Gilbert damping and nutation as successive, relativistic, corrections to the dynamics of a spin, in equilibrium with a quantum bath.

In the present paper we wish to explore the scenario, where a magnetic moment is in equilibrium with a vector bath, described by colored noise.
The correlation time of the noise sets the short--time scale, so the relativistic expansion of ref.~\cite{mondalRelativisticTheoryMagnetic2017} can be identified as the expansion in powers of the ratio of the correlation time to the period of precession.

We find that it is possible to recover both, Gilbert damping and nutation of the magnetic moment, as terms in such an expansion. 

It should be kept in mind that what is the ``most appropriate'' equation of motion (eom), that represents the motion of a collection of interacting magnetic moments, is still the subject of intense debate, that goes back a long time. 
Landau-Lifshitz~\cite{aharoniIntroductionTheoryFerromagnetism2000,*chikazumiPhysicsFerromagnetism1997} and Gilbert~\cite{gilbertAnomalousRotationalDamping1955} introduced an eom that described exclusively transverse damping, whereas Bloch considered an eom that described exclusively longitudinal damping for the coarse--grained (spatial) average of the magnetization of the interacting magnetic moments.
These equations have been extensively used to interpret measurements of spin relaxation and provide a phenomenological viewpoint both for the origin of the effective field that defines the precession axis for the average magnetization, as well as for the origin of the damping, whose effects can be reduced to a small number of {\it damping constants}. 
A lot of attempts have been made to provide a {\it microscopic} foundation for the equation of motion and, in particular, for accounting for the degrees of freedom that are behind the damping effects of the magnetization~\cite{seshadriDissipativeContributionsInternal1982,jayannavarBrownianMotionSpins1991,rossiDynamicsMagnetizationCoupled2005,vittoriaRelaxationMechanismOrdered2010}.
There have been many arguments about intrinsic and extrinsic effects, without, however, any insight into how these might be distinguished clearly in an invariant way. 

For example, some authors~\cite{seshadriDissipativeContributionsInternal1982,miyazakiBrownianMotionSpins1998} considered a phenomenological theory describing one classical spin, embedded in a medium, that acts as a bath.
This approach leads to the well-known Landau-Lifshitz equation (resp. Bloch equation) in several limiting cases, i.e. in the high temperature limit.
The origin of the bath, that describes the fluctuations of the average magnetization, was not spelled out, and only its role as an external {\it thermostat} (here called {\it fluctuostat} more generally) was assumed.
This approach highlights  that the damping is then a consequence of memory effects, i.e. non-local in time. 
Memory effects in their own right, were investigated theoretically in refs~\cite{boseRetardationEffectsLandauLifshitzGilbert2011}. 
Depending on the form of the memory kernel involved, it was found that these can lead to a compensation or even to an overcompensation of the damping, since called ``Gilbert damping''.

More recently and based on both a quantized spin and environment Hamiltonian, Anders {\it et.al} derive a general spin operator equation of motion that describes three-dimensional precession and damping and consistently account for effects arising from memory, coloured noise and quantum statistics~\cite{andersQuantumBrownianMotion2021}.
This reveals clearly resonant Lorentzian system--reservoir couplings that allow a systematic comparison of dynamics between Ohmic and non--Ohmic regimes.
The quantized spin+reservoir problem was also addressed before~\cite{nievesQuantumLandauLifshitzBlochEquation2014}, first in a attempt to justify the form taken by the classical Landau-Lifshitz-Bloch equation, without recognizing immediately the benefit of keeping, as long as possible, the memory kernel form induced by the motion equation of the quantum spin density operator.

In these approaches, it is implicitly assumed that the system does reach an equilibrium state, i.e. a state that is invariant under global time translations. 
How the system can, indeed, attain such a state has become a subject of considerable interest in the domain of glassy systems (for magnetic systems these are known as ``spin glasses''; an example of the vast literature is refs.~\cite{manivAntiferromagneticSwitchingDriven2021,bellettiNonequilibriumSpinGlassDynamics2008}). 
One way can be described as due to {\it inertia}, i.e. that the medium is not infinitely rigid~\cite{ashworthTransformationsInertialRotating1979}. 
This implies that the magnetic response depends not only on the magnetization itself, but on its velocity, as well, and, therefore, the equation of motion is of second order in time~\cite{rubiInertialEffectsNonequilibrium1999,ciorneiMagnetizationDynamicsInertial2011}.

The corresponding equations of motion can be identified with those of an Euler top, in a time--dependent external field, i.e. a torque. 
It is the impossibility of providing a local description of the dynamics in terms of one set of first order equations that leads to non--local effects. 
These can be captured by the so-called ``atomistic spin dynamics''--as implemented, for instance, by Bhattacharjee {\it et al.}~\cite{bhattacharjeeAtomisticSpinDynamic2012}. 
A particular motivation was of capturing processes in the femtosecond regime by including the  moment of inertia. 
They derived a generalized equation of motion for the magnetization dynamics in the semiclassical limit, which is non-local in both space and time. 
Consequently, they recovered a generalized Landau-Lifshitz-Gilbert equation, which includes the moment of inertia and a second derivative of the magnetization in time.

Going further with this idea, Pervishko {\it et al.}~\cite{pervishkoAnotherViewGilbert2018} proposed an alternative derivation of the Gilbert damping in a tensor form, within a mean-field approach. 
In this formalism, the itinerant electronic subsystem is considered in the presence of a nonequilibrium, classical magnetization field.
When this field is sufficiently smooth and slow on the scales determined by the mean free path and scattering rate of the conduction electrons, the induced nonlocal spin polarization can be approximated using a linear response Ansatz, thereby showing that the damping parameter emerges due to the coupling to the itinerant subsystem.
They derive a Kubo-Středa formula for the components of the Gilbert damping tensor and illustrate its relevance for the  two-dimensional Rashba ferromagnet, that can be realized at the interface between nonmagnetic and ferromagnetic layers. 
They argue that this approach can be further applied to identify properly the tensor structure of the Gilbert damping for more complicated model systems and real materials.

More recently, Mondal {\it et al.}~\cite{mondalGeneralisationGilbertDamping2018} identified the Gilbert damping and nutation terms as first- and second-order relativistic effects respectively, arising from the Foldy–-Wouthuysen transformation of a Dirac particle (that includes spin$-\frac{1}{2}$) motion under external fields, embedded in a material medium. 

What is particularly striking in all these approaches is that, while all end up with a description of  Gilbert damping and of the torque that drives nutation, they seem to  allow considerable ambiguity about the relative sign between the Gilbert damping and the nutation torque contribution.

While the microscopic origins of both Gilbert damping and magnetic inertia are still under debate, this uncertainty reflects a fundamental issue, that deserves closer scrutiny.

We wish to report on our efforts to resolve this ambiguity. We shall show that the coupling of a magnetic moment to a vector bath of colored noise is sufficient for describing the emergence of both Gilbert damping and nutation, along with the relative sign; in addition, it provides a well--defined route to equilibrium. 
The parameter that controls the relative significance of these effects is the ratio of the colored noise timescale to the precession period. 
This is where the vector nature of the bath is of relevance.

The plan of our paper is as follows:
In section~\ref{sec2} we describe our model for a magnetic moment in a vector bath.
In section~\ref{sec:numerical_results} we provide representative solutions of the equations of motion obtained by numerical integration and show how Gilbert damping and nutation can be unambiguously identified. 
In section~\ref{conclusion} we present our conclusions and ideas for further inquiry.  

\section{Magnetic moment in a bath}
\label{sec2}

Consider the {\em spatial} average of the magnetization ${\bm M}$ of a block of magnetic material.
The ``reduced'' magnetization, $\bm{m}\equiv{\bm M}/M_s$, depends only on time and its dynamics can be described by its precession about an effective field, which can be written as the sum of two vectors $\bm{\omega}_0(t)+\delta\bm{\omega}(t)$.
$\bm{\omega}_0(t)$ is defined, in turn, as the sum of the external magnetic field, applied on the magnetic system, and of the magnetic field, produced by the average magnetization of the surrounding medium, i.e. the reaction field.

$\delta\bm{\omega}(t)$ is a stochastic field, and is characterized phenomenologically by a single relaxation time $\tau$.
It describes the fluctuations of the magnetic response of the medium, in which the magnetic block is found. 

We can describe the equilibrium of the magnetic block with the medium, by the statement that $\langle\delta{\bm\omega}\rangle\equiv\gamma\mu_0M_s\chi^{-1}\langle{\bm m}\rangle=\Omega_s\chi^{-1}\langle{\bm m}\rangle$, where $\chi$ is the susceptibility (not a function of time), and $\gamma$ is the gyromagnetic ratio.
Here the average is taken over the realizations of the surrounding medium, considered as a bath.
This statement means that the expectation value of the fluctuating field at equilibrium is aligned with and proportional to the expectation value of the magnetization~\cite{neelMagnetismLocalMolecular1971}.
When ${\chi},$ which is identified as the cumulant of the spin-spin function, depends explicitly on time, a convolution between the fluctuating field and the magnetization has to be used~\cite{guimaraesComparativeStudyMethodologies2019}.

This procedure focuses on the "relevant" degrees of freedom, labelled by ${\bm m}$ and sets them apart from the "irrelevant" variables, labelled by $\delta{\bm\omega}$.
We are not interested, in the following, in the microscopic mechanisms that may produce  the effects of these variables~\cite{kamberskyLandauLifshitzRelaxation1970,zwanzigNonequilibriumStatisticalMechanics2001}, just on their collective dynamics on the ``relevant'' degrees of freedom. 
(It has been proposed~\cite{parisiSupersymmetricFieldTheories1982,nicolisParticleEquilibriumBath2019} that the symmetry that expresses the property that the physics should not depend on how the ``dynamical'' from the degrees of freedom, that can define the ``bath'', are chosen, is supersymmetry.)

These considerations can be expressed mathematically as follows:
\begin{align}
    \label{transversem}
    \frac{d{\bm m}}{dt}&=\left({\bm\omega}_0+\delta{\bm\omega}\right)\times{\bm m}\\
    \label{fluctuation}
    \frac{d\delta{\bm\omega}}{dt}&=-\frac{1}{\tau}\left(\delta{\bm\omega}-\Omega_s\chi^{-1}{\bm m}\right)+\Omega_s{\bm\eta}
\end{align}
where $\bm{\eta}$ is a random field, with ultra--local Gaussian correlations,  that describes the bath, which will be taken as thermal, in what follows, concretely:
\begin{align}
    \langle{\eta}_I(t)\rangle&=0\label{prop1eta}\\
    \langle{\eta}_I(t){\eta}_J(t')\rangle&=2D\delta_{IJ}\delta(t-t')\label{prop2eta}
\end{align}
where $I,J$ are the indices of the vector components.
$D$ is the amplitude  of the noise and provides the definition of the  temperature $T$, through the Boltzmann--Einstein relation, $D\propto k_\mathrm{B}T/\hbar$, thereby expressing the fluctuation-dissipation theorem, for the bath. 
That the temperature is well--defined is ensured by the property that the noise field $\bm{\eta}(t)$ is drawn from a stationary stochastic process, i.e. enjoys {\em global} time translation invariance.
Equations~\eqref{transversem} and \eqref{fluctuation} were first defined  in~\cite{miyazakiBrownianMotionSpins1998} and evaluated in atomistic spin simulations~\cite{atxitiaUltrafastSpinDynamics2009}. 
It should be stressed that this does not imply that the 2--point function of the magnetic moment will have a simple dependence on the temperature, due to the fact that its fluctuations, generically, will not be Gaussian~\cite{zinn-justinPhaseTransitionsRenormalization2007}. 

Eq.~\eqref{transversem} is purely transverse, therefore, the norm of ${\bm m}$ is conserved, if ${\bm m}\cdot\dot{\bm m}=0\Leftrightarrow (d/dt)(||{\bf m}||^2)=0$.
The latter relation is, of course, true, in the absence of the bath; it does require, however, another definition in its presence, since the derivative is a singular quantity~\cite{zinn-justinQuantumFieldTheory2002,cugliandoloRulesCalculusPath2017}.
Such a definition can be obtained from the so-called  Schwinger--Dyson identities \cite{zinn-justinPathIntegralsQuantum2009}, namely as 
\begin{equation}
    \label{SchwingerDyson}
    \begin{array}{l}
    \displaystyle
    \left\langle\bm{m}\cdot\frac{d\bm{m}}{dt}\right\rangle=0.\\
    \end{array}
\end{equation} 

The field $\delta{\bm\omega}$ is defined by the stochastic differential equation (SDE)~in eq.~\eqref{fluctuation}. 
Its solution can be shown to be an Ornstein-Uhlenbeck process~\cite{gardinerStochasticMethodsHandbook2009}.
Therefore, ${\bm m}(t)$ becomes a stochastic process, as well; moreover, the noise, that enters additively in the equation for $\delta\bm{\omega}$, becomes multiplicative for ${\bm m}(t)$; which implies that its correlation functions acquire a non--trivial dependence on the temperature, defined through the bath.
This is, often described as a ``breakdown'' of the fluctuation--dissipation theorems~\cite{vankampenStochasticProcessesPhysics1992,tranchidaHierarchiesLandauLifshitzBlochEquations2018}.
However, what this, simply, means is that the non-linearities induce a non--trivial, but quite transparent, dependence of the noise on the dynamics of the magnetization; the two are, just, intertwined in a way that is more subtle than hitherto acknowledged.
Indeed, this can be understood in terms of the variables that can resolve the dynamics of the bath, as an expression of reparametrization invariance in the space of fields. 

First, suppose for simplicity that the system is not in contact with the bath; ${\bm\eta}(t)$ is absent from eq.~(\ref{fluctuation}).
Then, eqs.(\ref{transversem},\ref{fluctuation}) define the dynamics of a deterministic system and can be explicitly solved:
First of all, equation \eqref{fluctuation} can be solved for $\delta{\bm\omega}$ in terms of $\bm{m}(t)$:
\begin{equation}
    \begin{aligned}
    \delta{\bm\omega}&=\frac{\Omega_s\chi^{-1}}{\tau}\int_{-\infty}^{t}e^{-\frac{t-t'}{\tau}}{\bm m}(t')dt'\\
    &=\frac{\Omega_s\chi^{-1}}{\tau}\int_0^\infty e^{-\frac{u}{\tau}}{\bm m}(t-u)du
    \label{solutionfluctuation}
    \end{aligned}
\end{equation}
The equation~\eqref{solutionfluctuation} can then be introduced in eq.~\eqref{transversem} to produce an integral-differential equation for ${\bm m}$: 
\begin{equation}
    \frac{d{\bm m}}{dt}=\left({\bm\omega}_0+\frac{\Omega_s\chi^{-1}}{\tau}\int_0^\infty e^{-\frac{u}{\tau}}{\bm m}(t-u)du\right)\times{\bm m}
\end{equation}
The integral highlights the dependence of the solution on the full history of the magnetization, prior to time $t$, as well as the putative effects of the damping induced by the memory kernel with a characteristic time $\tau$, defined by eq.~\eqref{fluctuation}.
Indeed one of the purposes of this paper is to provide an intrinsic definition of such damping effects in an invariant way. 

In order to find approximate solutions, it is useful to expand ${\bm m}$ in a Taylor series about some reference time $t$ and exchange the sum and the integral. 
Assuming that Fubini's theorem holds~\cite{weirLebesgueIntegrationMeasure1973}, we thus find 
\begin{equation}
    \begin{aligned}
    \delta{\bm\omega}&=\frac{\Omega_s\chi^{-1}}{\tau}\sum_{n=0}^\infty\frac{(-1)^n}{n!}\frac{d^n{\bm m}}{dt^n}\int_0^\infty e^{-\frac{u}{\tau}}u^ndu\\
    &=\Omega_s\chi^{-1}\sum_{n=0}^\infty\frac{(-1)^n}{n!}\frac{d^n{\bm m}}{dt^n}\tau^n\Gamma(1+n)\\
    &=\Omega_s\chi^{-1}\sum_{n=0}^\infty(-\tau)^n\frac{d^n{\bm m}}{dt^n}.
    \end{aligned}
    \label{deltaomega}
\end{equation}
Of course, it is by no means obvious either that this series converges, or that it is even legitimate to exchange sum and integral; we shall try to provide {\em a posteriori} checks that are sensitive to these issues. 

We shall now try to interpret the properties of the magnetization, that are sensitive to our truncating the series at a given order.
When the sum stops at $n=1$, eq.\eqref{transversem} takes the form
\begin{equation}
    \begin{aligned}
        \frac{d{\bm m}}{dt}&\approx\left({\bm\omega}_0+\Omega_s\chi^{-1}{\bm m}-\Omega_s\chi^{-1}\tau\frac{d{\bm m}}{dt}\right)\times{\bm m}\\
        &={\bm\omega}_0\times{\bm m}+\alpha{\bm m}\times\frac{d{\bm m}}{dt}
    \end{aligned}
    \label{approx1}
\end{equation}
where $\alpha\equiv\Omega_s\tau\chi^{-1}$ can be, therefore, identified as the Gilbert damping constant, and eq.~\eqref{approx1} is the eom written in the standard Gilbert form~\cite{gilbertAnomalousRotationalDamping1955}.
This expression for $\alpha$ appears consistent with other forms reported in the literature~\cite{garateGilbertDampingConducting2009,guimaraesComparativeStudyMethodologies2019}.
It is therefore not surprising that eventually the tensor character of both the inverse of the susceptibility $\chi$ and the relaxation time $\tau$ produces a tensor damping parameter $\alpha$, a feature already reported in ferromagnetic metals assuming a torque-torque correlation model~\cite{garateNonadiabaticSpintransferTorque2009,thonigNonlocalGilbertDamping2018,hickeyOriginIntrinsicGilbert2009} in the highly anisotropic scattering regime of magnons.

Upon including the $n=2$ term, the equation for the magnetization takes the form

\begin{equation}
        \begin{aligned}
        \frac{d{\bm m}}{dt}\approx&\left({\bm\omega}_0+\Omega_s\chi^{-1}{\bm m}-\Omega_s\chi^{-1}\tau\frac{d{\bm m}}{dt}\right)\times{\bm m}\\
        &+\Omega_s\chi^{-1}\tau^2\frac{d^2{\bm m}}{dt^2}\times{\bm m}\\
        =&{\bm\omega}_0\times{\bm m}+\alpha{\bm m}\times\left(\frac{d{\bm m}}{dt}-\tau\frac{d^2{\bm m}}{dt^2}\right).
        \end{aligned}
    \label{approx2}
\end{equation}

The term proportional to $\bm{m}\times (d^2\bm{m}/dt^2)$ can be interpreted as describing the "nutation" of the magnetization~\cite{fahnleGeneralizedGilbertEquation2011,*fahnleErratumGeneralizedGilbert2013}. 
It should be noted, at this point that this is the first term that is, manifestly, symmetric under time--reversal.
An issue of considerable interest is that of the relative sign of the coefficients of the terms in the equation of motion.
Let us note that the sign of the inertial damping (last term) seems to be opposite to the sign of the usual damping term (second term), which is in agreement with the theory of dampened magnetostriction, first introduced by Suhl~\cite{suhlTheoryMagneticDamping1998,*suhlRelaxationProcessesMicromagnetics2007}.
This is in contrast with reference~\cite{ciorneiMagnetizationDynamicsInertial2011}, where the signs of the two damping terms are the same.
However the microscopic description in the two cases is completely different.

What we have thus shown is that both, the Gilbert damping and the nutation  term can be deduced  as the consequence of the coupling of a magnetic moment to an external field, upon taking into account the coupling to the bath self--consistently. 

This constitutes the central result of the paper.

In this deterministic situation and because $\tau>0$, the divergence of the volume of the phase space is negative--it shrinks, due to dissipation. 
At this level of truncation the existence of an equilibrium state for the magnetization that is unique and is described by a point is obvious. 
What is by no means obvious is what happens when the non--local effects, described by the higher order terms, are taken into account. 

Setting this issue aside, for the moment, let us now take into account the bath, at this approximation.

When the noise field ${\bm\eta}$ is present, $\delta{\bm\omega}$ becomes a stochastic field, which contains an extra term. 
This term takes into account the noise field in the memory kernel as follows:
\begin{equation}
    \frac{\delta{\bm\omega}}{\Omega_s}=\chi^{-1}\sum_{n=0}^\infty(-\tau)^n\frac{d^n{\bm m}}{dt^n}+\int_{0}^\infty e^{-\frac{u}{\tau}}{\bm\eta}(t-u)du
\end{equation}
Let us call ${\bm\Omega}(t)\equiv\int_0^\infty e^{-\frac{u}{\tau}}{\bm\eta}(t-u)du$, the extra stochastic field.
Equations \eqref{prop1eta},\eqref{prop2eta} imply that the random field ${\bm\Omega}$ has the following properties:
\begin{align}
    \langle{\Omega}_I(t)\rangle&=0\label{prop1omega}\\
    \langle{\Omega}_I(t){\Omega}_J(t')\rangle&=D\tau\delta_{IJ}e^{-\frac{|t-t'|}{\tau}}\label{prop2omega}
\end{align}
which means that it describes colored noise!
This implies, in turn, for the magnetization that its correlation functions are, generically, those of a centered and colored noise stochastic process, and not of a white noise process, as it is usually assumed.
We shall show now show that the approximations involved in the truncation to second order, i.e. including the nutation term, are self--consistent by solving the equations~(\ref{transversem}) numerically. 
\section{Numerical results}
\label{sec:numerical_results}

In order to check on how the signature of  the Gilbert damping and that of  nutation,  produced by the fluctuating field ${\delta\bm\omega}$ is imprinted in the magnetization profile, we solve the coupled equations~(\ref{transversem}) numerically. 

Precession can be readily identified as the rotation of the magnetization around a given axis.

Nutation is the additional effect produced on the magnetization by the motion of this axis with time. 

Upon averaging over the realizations of the noise, if $\langle{\bm m}\rangle$ is a constant vector at equilibrium, then ${\langle{\delta{\bm\omega}}\rangle}$ also becomes a constant vector, proportional to $\langle{\bm m}\rangle$.
That means that the magnetization spins first around ${\bm\omega}_0$ at short times and then settles to spinning around ${\bm\omega}_0+\langle\delta{\bm\omega}\rangle$ at long times.
But the torque produced at that time is ${\bm\omega}_0\times\langle{\bm m}\rangle$, because of the proportionality between $\langle\delta{\bm\omega}\rangle$ and ${\langle{\bm m}\rangle}$.
As a consequence, only during a transient time, when $\delta{\bm\omega}$ strongly varies, can  the motion of the average magnetization be strongly affected. 

\begin{figure}[htb]
    \begin{center}
        \includegraphics[width=\columnwidth]{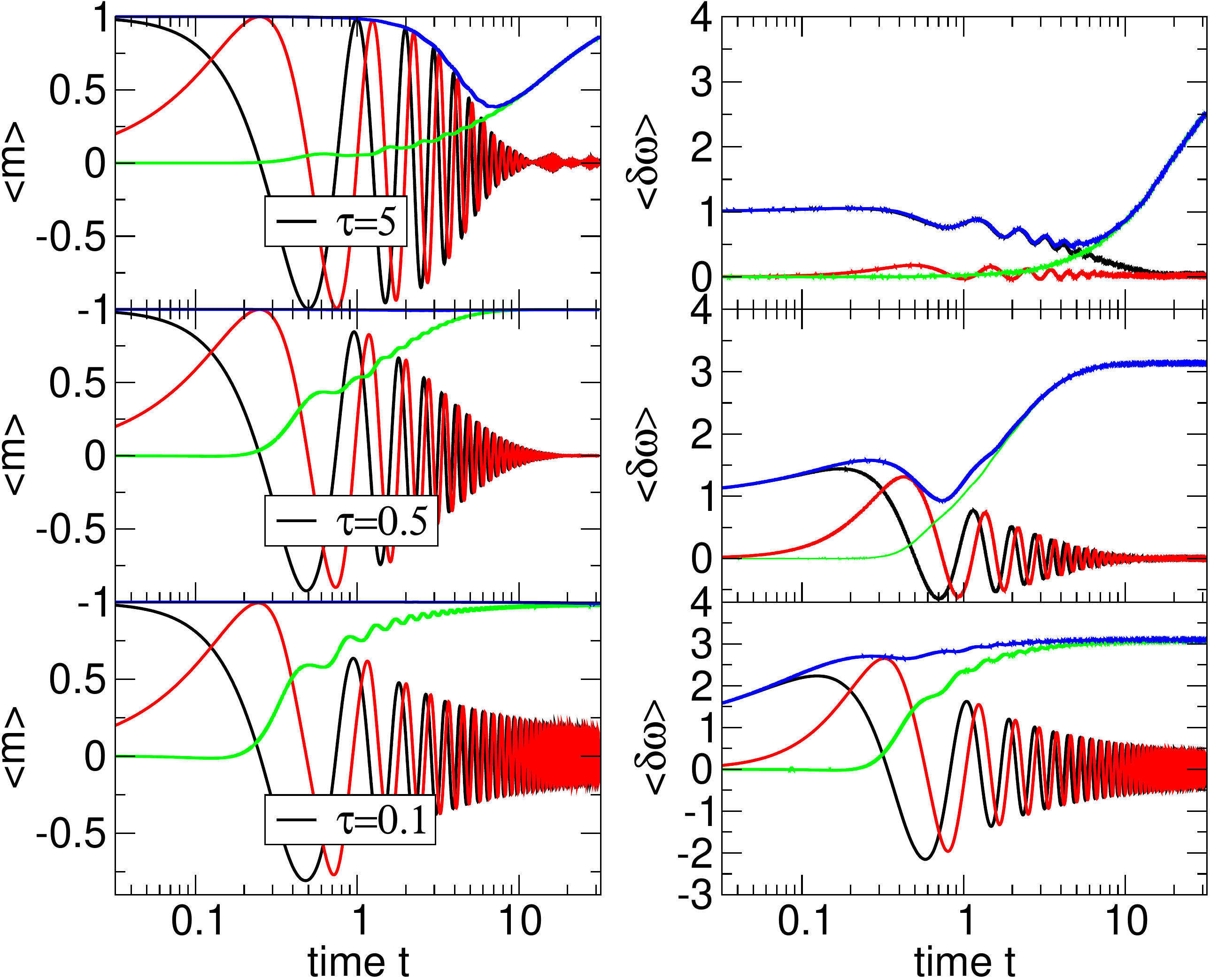}
    \end{center}
    \caption{(color online) Dynamics of the average magnetization (left panels) and fluctuation field (right panels) for a varying $\tau$ parameter. Conditions are $\bm{\omega}_0=<0,0,2\pi>$, $D=50$, $\Omega_s\chi^{-1}=\pi$, $\bm{m}(0)=<1,0,0>$, ${\delta{\bm\omega}}(0)=<1,0,0>$. Components of $x$, $y$, $z$ are in black, red and green respectively. The norm is displayed in blue.}
    \label{fig1}
\end{figure}

All these features can be read off figure~{\ref{fig1}}, that displays both the motion of the average ${\bm m}$ and ${\delta{\bm\omega}}$ taken over more than 1000 realizations of the noise, and for different values of the correlation time of the noise, $\tau$.

The equations~\eqref{transversem} are integrated globally with an explicit 4$^\mathrm{th}$ order Runge-Kutta algorithm and a variable stepping scheme, with only a renormalization of the magnetization at each step, in order to produce a precession and nutation motion consistent on the $S^2$ sphere.
Better symplectic algorithms~\cite{omelyanSymplecticAnalyticallyIntegrable2003}, that preserve the structure of the equations of motion, can be used, but they do not affect the conclusions drawn.

What we observe here is that the average magnetization $\langle{\bm m}\rangle$ tends to align with  the effective field along the $z$-axis, by producing a dampened motion and a wriggling movement of the $\langle m_z\rangle$ component, which is characteristic of a high frequency nutation effect, because of the finite values of $\tau$ and $\Omega_s\chi^{-1}$.
When the susceptibility $\chi$ is decreased, while keeping all the other parameters fixed, the internal precession field, coming from the fluctuations, dominates the natural precession field ${\bm{\omega}_0}$, that increases the precession pulsation.
We observe that increasing $\tau$ and reducing $M_s$ does indeed enhance the effects of the nutation term.
Moreover when $\tau$ is large, the diffusive term, that is generated by the noise, dominates the motion of the magnetization. 
A consequence is that $\langle{\bm m}(t)\rangle$ cannot stay constant even if, for all values of $\tau$, $\langle {\bm m}.{\bm m}\rangle=1$ by construction.
When the time is long enough to capture the growing main component of the magnetization then $\langle{\bm m}\rangle$ aligns itself on $\langle{\delta\bm{\omega}}\rangle$. 
When $\tau$ is small, in the transient regime, the fluctuating field $\langle{\delta\bm{\omega}}\rangle$ cannot be sufficiently dampened and follows more closely the dynamics of the magnetization. 
For low values of the noise amplitude, the dynamics of the average $\langle{\delta\bm{\omega}}\rangle$ is insensitive to the noise amplitude and its leading motion is described by $\tau\langle{\delta{\bm{\omega}}}\rangle\approx \alpha\langle{{\bm m}}\rangle$. 
When $\tau$ takes values of $O(1/\omega_0)$, the leading motion of $\langle{\bm m}\rangle$ is given by the Gilbert equation of precession 
$d\langle{\bm m}\rangle/dt\approx\left({\bm{\omega}_0-\alpha\langle d{\bm m}}\rangle/dt\right)\times \langle{\bm m}\rangle$, that produces in return a dampened motion of $\langle\delta{\bm\omega}\rangle$.

One conclusion of this study is identifying the appropriate dimensionless combinations.
Our results motivate defining the dimensionless quantities ${\bm X}_0\equiv\tau{\bm{\omega}_0}$ and ${\bm X}\equiv\tau\delta{\bm{\omega}}$. 
In terms of these te equations of motion~\ref{transversem} and \ref{fluctuation} take the form (upon defining $x=t/\tau$)
\begin{equation}
\begin{aligned}
    \frac{d\bm{m}}{dx}&=({\bm X}_0+{\bm{X}})\times{\bm m}\\
    \frac{d\bm{X}}{dx}&=-({\bm X}-\alpha{\bm m})+\Delta{\bm\eta}
\end{aligned}
\end{equation}
where $\Delta\equiv\Omega_s\tau$ and $\langle\eta_i(x)\eta_j(x')\rangle=2D/\tau\delta_{ij}\delta(x-x')$.
In the particular case  where ${\bm X}_0={\bm 0}$, i.e. when no external torque acts on the magnetization and with $\alpha\neq 0$, then $\langle\bm{m}\rangle\equiv{\bm m}(0)$ is a constant of motion. 

Upon averaging over the noise realizations, the dynamics of $\langle{\bm X}\rangle$ is given by $\langle{\bm X}\rangle=({\bm{X}_0}-\alpha{\bm m}(0))e^{-x}+\alpha{\bm m}(0)$.
Thus the fluctuating field at equilibrium is given by $\langle{\bm X}\rangle_\infty=\alpha{\bm m}(0)$ and no torque acts on the magnetization, keeping it  constant over time.

\begin{figure}[htb!]
    \begin{center}
        \includegraphics[width=\columnwidth]{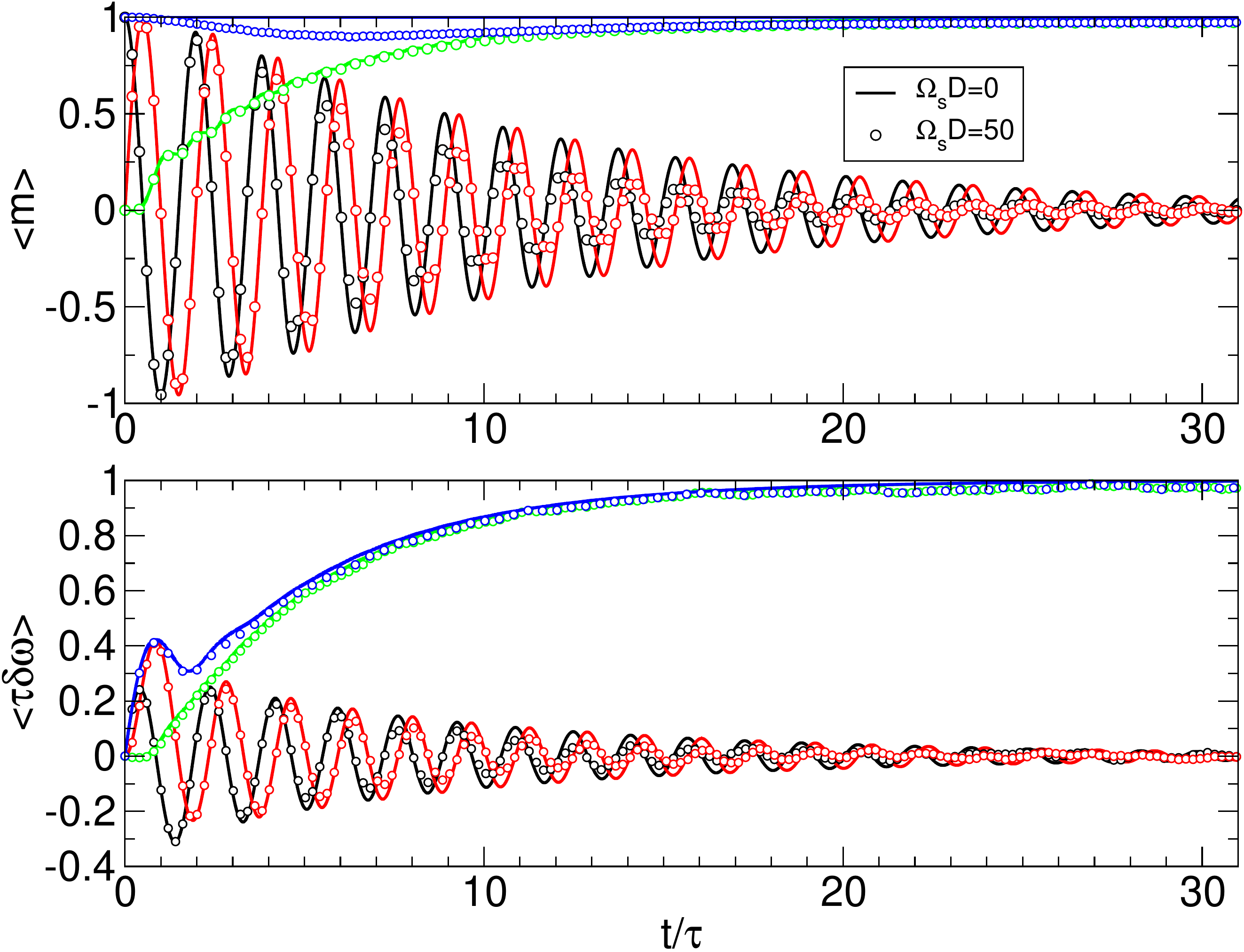}
    \end{center}
    \caption{(color online) Dynamics of the average magnetization (up panel) and fluctuation field (down) for a varying $\Omega_sD$ parameter. Conditions are $\tau\bm{\omega}_0=<0,0,\pi>$, $\alpha=1$, $\bm{m}(0)=<1,0,0>$, ${\tau\delta{\bm\omega}}(0)=<0,0,0>$. Components of $x$, $y$, $z$ are in black, red and green respectively. The norm is displayed in blue.}
    \label{fig2}
\end{figure}

The figure~\ref{fig2} displays the dampened motion of ${\bm m}$ and ${\bm X}$ as a function of the dimensionless time, for two configurations~: $\Omega_sD=0$, i.e. without thermal noise, and $\Omega_sD=50$, for the same external field $\bm{X}_0=<0,0,\pi>$.
The longitudinal behavior of the average magnetization is clearly visible by the decrease of the average norm $\|\langle{\bm m}\rangle\|$.
Moreover because the average eom for ${\bm X}$ is independent of the noise amplitude, as depicted, this is not the case for the average magnetization, because there $\langle {\bm X}\times {\bm m}\rangle\neq\langle {\bm X}\rangle\times \langle{\bm m}\rangle$.    


\section{Conclusions and outlook}
\label{conclusion}

In this paper, we have shown that the mechanism of Gilbert damping of the precession, as well as the effects of nutation can be understood in terms of an effective interaction between magnetic moments and the fluctuations of their effective fields, when the latter are described by colored noise in a systematic expansion in powers of the ratio of the correlation time of the noise to the period defined by the precession torque. 

We have identified a relation between the Gilbert damping parameter and the static (or spectral) inverse susceptibility of the material, with the contribution to a characteristic relaxation time, that can be assigned to magnon scattering mechanisms, in the relaxation time approximation.

It is stressed that if it were possible to perform measurements that could resolve the contribution of the nutation loops, as they are superimposed on the usual precession motion of the magnetic moments, it would be possible to find which processes provide the dominant contribution leading to inertial damping, as recently been reported~\cite{neerajInertialSpinDynamics2020}.

The relative sign between the Gilbert damping and the inertia term is {\em negative} as a consequence of the fact that these two terms represent successive contributions of the Taylor expansion. 
Therefore studies that assume that these terms have the same sign make additional assumptions, that it would be very interesting to spell out.

The results obtained here relied on the equations of motion alone.
To better understand the space of states of the magnetization, it will be useful to adapt the techniques used in ref.~\cite{tranchidaHierarchiesLandauLifshitzBlochEquations2018} and to to understand the microscopic degrees of freedom that can define the bath in an invariant way it is necessary to implement the program that is sketched in ref.~\cite{nicolisFinitedimensionalColoredFluctuationdissipation2017}.

Since the magnetization vector naturally evolves according to Nambu mechanics, it will, also, be interesting to understand how Nambu mechanics may accommodate Gilbert damping and nutation. 
Gilbert damping has been studied, in this context, already, using different tools, in refs.~\cite{thibaudeauNambuMechanicsStochastic2017}.

Of course probing how the truncation to $n=2$ breaks down and how it may be completed remains to be understood.

We hope to report on progress on these issues in future work.


\begin{thebibliography}{49}%
    \makeatletter
    \providecommand \@ifxundefined [1]{%
     \@ifx{#1\undefined}
    }%
    \providecommand \@ifnum [1]{%
     \ifnum #1\expandafter \@firstoftwo
     \else \expandafter \@secondoftwo
     \fi
    }%
    \providecommand \@ifx [1]{%
     \ifx #1\expandafter \@firstoftwo
     \else \expandafter \@secondoftwo
     \fi
    }%
    \providecommand \natexlab [1]{#1}%
    \providecommand \enquote  [1]{``#1''}%
    \providecommand \bibnamefont  [1]{#1}%
    \providecommand \bibfnamefont [1]{#1}%
    \providecommand \citenamefont [1]{#1}%
    \providecommand \href@noop [0]{\@secondoftwo}%
    \providecommand \href [0]{\begingroup \@sanitize@url \@href}%
    \providecommand \@href[1]{\@@startlink{#1}\@@href}%
    \providecommand \@@href[1]{\endgroup#1\@@endlink}%
    \providecommand \@sanitize@url [0]{\catcode `\\12\catcode `\$12\catcode
      `\&12\catcode `\#12\catcode `\^12\catcode `\_12\catcode `\%12\relax}%
    \providecommand \@@startlink[1]{}%
    \providecommand \@@endlink[0]{}%
    \providecommand \url  [0]{\begingroup\@sanitize@url \@url }%
    \providecommand \@url [1]{\endgroup\@href {#1}{\urlprefix }}%
    \providecommand \urlprefix  [0]{URL }%
    \providecommand \Eprint [0]{\href }%
    \providecommand \doibase [0]{https://doi.org/}%
    \providecommand \selectlanguage [0]{\@gobble}%
    \providecommand \bibinfo  [0]{\@secondoftwo}%
    \providecommand \bibfield  [0]{\@secondoftwo}%
    \providecommand \translation [1]{[#1]}%
    \providecommand \BibitemOpen [0]{}%
    \providecommand \bibitemStop [0]{}%
    \providecommand \bibitemNoStop [0]{.\EOS\space}%
    \providecommand \EOS [0]{\spacefactor3000\relax}%
    \providecommand \BibitemShut  [1]{\csname bibitem#1\endcsname}%
    \let\auto@bib@innerbib\@empty
    \bibitem [{\citenamefont {Neeraj}\ \emph {et~al.}(2020)\citenamefont {Neeraj},
      \citenamefont {Awari}, \citenamefont {Kovalev}, \citenamefont {Polley},
      \citenamefont {Zhou~Hagstr{\"o}m}, \citenamefont {Arekapudi}, \citenamefont
      {Semisalova}, \citenamefont {Lenz}, \citenamefont {Green}, \citenamefont
      {Deinert}, \citenamefont {Ilyakov}, \citenamefont {Chen}, \citenamefont
      {Bawatna}, \citenamefont {Scalera}, \citenamefont {{d'Aquino}}, \citenamefont
      {Serpico}, \citenamefont {Hellwig}, \citenamefont {Wegrowe}, \citenamefont
      {Gensch},\ and\ \citenamefont {Bonetti}}]{neerajInertialSpinDynamics2020}%
      \BibitemOpen
      \bibfield  {author} {\bibinfo {author} {\bibfnamefont {K.}~\bibnamefont
      {Neeraj}}, \bibinfo {author} {\bibfnamefont {N.}~\bibnamefont {Awari}},
      \bibinfo {author} {\bibfnamefont {S.}~\bibnamefont {Kovalev}}, \bibinfo
      {author} {\bibfnamefont {D.}~\bibnamefont {Polley}}, \bibinfo {author}
      {\bibfnamefont {N.}~\bibnamefont {Zhou~Hagstr{\"o}m}}, \bibinfo {author}
      {\bibfnamefont {S.~S. P.~K.}\ \bibnamefont {Arekapudi}}, \bibinfo {author}
      {\bibfnamefont {A.}~\bibnamefont {Semisalova}}, \bibinfo {author}
      {\bibfnamefont {K.}~\bibnamefont {Lenz}}, \bibinfo {author} {\bibfnamefont
      {B.}~\bibnamefont {Green}}, \bibinfo {author} {\bibfnamefont {J.-C.}\
      \bibnamefont {Deinert}}, \bibinfo {author} {\bibfnamefont {I.}~\bibnamefont
      {Ilyakov}}, \bibinfo {author} {\bibfnamefont {M.}~\bibnamefont {Chen}},
      \bibinfo {author} {\bibfnamefont {M.}~\bibnamefont {Bawatna}}, \bibinfo
      {author} {\bibfnamefont {V.}~\bibnamefont {Scalera}}, \bibinfo {author}
      {\bibfnamefont {M.}~\bibnamefont {{d'Aquino}}}, \bibinfo {author}
      {\bibfnamefont {C.}~\bibnamefont {Serpico}}, \bibinfo {author} {\bibfnamefont
      {O.}~\bibnamefont {Hellwig}}, \bibinfo {author} {\bibfnamefont {J.-E.}\
      \bibnamefont {Wegrowe}}, \bibinfo {author} {\bibfnamefont {M.}~\bibnamefont
      {Gensch}},\ and\ \bibinfo {author} {\bibfnamefont {S.}~\bibnamefont
      {Bonetti}},\ }\href {https://doi.org/10.1038/s41567-020-01040-y} {\bibfield
      {journal} {\bibinfo  {journal} {Nature Physics}\ ,\ \bibinfo {pages} {1}}
      (\bibinfo {year} {2020})}\BibitemShut {NoStop}%
    \bibitem [{\citenamefont {Gilbert}\ and\ \citenamefont
      {Kelly}(1955)}]{gilbertAnomalousRotationalDamping1955}%
      \BibitemOpen
      \bibfield  {author} {\bibinfo {author} {\bibfnamefont {T.~L.}\ \bibnamefont
      {Gilbert}}\ and\ \bibinfo {author} {\bibfnamefont {J.~M.}\ \bibnamefont
      {Kelly}},\ }in\ \href@noop {} {{\selectlanguage {English}\emph {\bibinfo
      {booktitle} {Conference on {{Magnetism}} and {{Magnetic Materials}}}}}}\
      (\bibinfo {address} {{Pittsburgh, PA}},\ \bibinfo {year} {1955})\ pp.\
      \bibinfo {pages} {253--263}\BibitemShut {NoStop}%
    \bibitem [{\citenamefont {Makhfudz}\ \emph {et~al.}(2020)\citenamefont
      {Makhfudz}, \citenamefont {Olive},\ and\ \citenamefont
      {Nicolis}}]{makhfudzNutationWavePlatform2020}%
      \BibitemOpen
      \bibfield  {author} {\bibinfo {author} {\bibfnamefont {I.}~\bibnamefont
      {Makhfudz}}, \bibinfo {author} {\bibfnamefont {E.}~\bibnamefont {Olive}},\
      and\ \bibinfo {author} {\bibfnamefont {S.}~\bibnamefont {Nicolis}},\ }\href
      {https://doi.org/10.1063/5.0013062} {\bibfield  {journal} {\bibinfo
      {journal} {Appl. Phys. Lett.}\ }\textbf {\bibinfo {volume} {117}},\ \bibinfo
      {pages} {132403} (\bibinfo {year} {2020})}\BibitemShut {NoStop}%
    \bibitem [{\citenamefont {{Cohen-Tannoudji}}\ \emph {et~al.}(1993)\citenamefont
      {{Cohen-Tannoudji}}, \citenamefont {Diu}, \citenamefont {Lalo{\"e}},\ and\
      \citenamefont {Hemley}}]{cohen-tannoudjiQuantumMechanics1993}%
      \BibitemOpen
      \bibfield  {author} {\bibinfo {author} {\bibfnamefont {C.}~\bibnamefont
      {{Cohen-Tannoudji}}}, \bibinfo {author} {\bibfnamefont {B.}~\bibnamefont
      {Diu}}, \bibinfo {author} {\bibfnamefont {F.}~\bibnamefont {Lalo{\"e}}},\
      and\ \bibinfo {author} {\bibfnamefont {S.~R.}\ \bibnamefont {Hemley}},\
      }\href@noop {} {\emph {\bibinfo {title} {Quantum Mechanics}}},\ A
      {{Wiley}}-{{Interscience}} Publication\ (\bibinfo  {publisher} {{Wiley
      [u.a.]}},\ \bibinfo {address} {{New York, NY}},\ \bibinfo {year}
      {1993})\BibitemShut {NoStop}%
    \bibitem [{\citenamefont {Thibaudeau}\ \emph {et~al.}(2017)\citenamefont
      {Thibaudeau}, \citenamefont {Nussle},\ and\ \citenamefont
      {Nicolis}}]{thibaudeauNambuMechanicsStochastic2017}%
      \BibitemOpen
      \bibfield  {author} {\bibinfo {author} {\bibfnamefont {P.}~\bibnamefont
      {Thibaudeau}}, \bibinfo {author} {\bibfnamefont {T.}~\bibnamefont {Nussle}},\
      and\ \bibinfo {author} {\bibfnamefont {S.}~\bibnamefont {Nicolis}},\ }\href
      {https://doi.org/10.1016/j.jmmm.2017.01.088} {\bibfield  {journal} {\bibinfo
      {journal} {J. Magn. Magn. Mater.}\ }\textbf {\bibinfo {volume} {432}},\
      \bibinfo {pages} {175} (\bibinfo {year} {2017})}\BibitemShut {NoStop}%
    \bibitem [{\citenamefont {Mondal}\ \emph {et~al.}(2017)\citenamefont {Mondal},
      \citenamefont {Berritta}, \citenamefont {Nandy},\ and\ \citenamefont
      {Oppeneer}}]{mondalRelativisticTheoryMagnetic2017}%
      \BibitemOpen
      \bibfield  {author} {\bibinfo {author} {\bibfnamefont {R.}~\bibnamefont
      {Mondal}}, \bibinfo {author} {\bibfnamefont {M.}~\bibnamefont {Berritta}},
      \bibinfo {author} {\bibfnamefont {A.~K.}\ \bibnamefont {Nandy}},\ and\
      \bibinfo {author} {\bibfnamefont {P.~M.}\ \bibnamefont {Oppeneer}},\ }\href
      {https://doi.org/10.1103/PhysRevB.96.024425} {\bibfield  {journal} {\bibinfo
      {journal} {Phys. Rev. B}\ }\textbf {\bibinfo {volume} {96}},\ \bibinfo
      {pages} {024425} (\bibinfo {year} {2017})},\ \Eprint
      {https://arxiv.org/abs/1704.01559} {arXiv:1704.01559} \BibitemShut {NoStop}%
    \bibitem [{\citenamefont
      {Aharoni}(2000)}]{aharoniIntroductionTheoryFerromagnetism2000}%
      \BibitemOpen
      \bibfield  {author} {\bibinfo {author} {\bibfnamefont {A.}~\bibnamefont
      {Aharoni}},\ }\href@noop {} {{\selectlanguage {English}\emph {\bibinfo
      {title} {Introduction to the {{Theory}} of {{Ferromagnetism}}}}}}\ (\bibinfo
      {publisher} {{Clarendon Press}},\ \bibinfo {year} {2000})\BibitemShut
      {NoStop}%
    \bibitem [{\citenamefont
      {Chikazumi}(1997)}]{chikazumiPhysicsFerromagnetism1997}%
      \BibitemOpen
      \bibfield  {author} {\bibinfo {author} {\bibfnamefont {S.}~\bibnamefont
      {Chikazumi}},\ }\href@noop {} {{\selectlanguage {English}\emph {\bibinfo
      {title} {Physics of {{Ferromagnetism}}}}}},\ International {{Series}} of
      {{Monographs}} on {{Physics}}\ (\bibinfo  {publisher} {{Oxford Science
      Publications}},\ \bibinfo {year} {1997})\BibitemShut {NoStop}%
    \bibitem [{\citenamefont {Seshadri}\ and\ \citenamefont
      {Lindenberg}(1982)}]{seshadriDissipativeContributionsInternal1982}%
      \BibitemOpen
      \bibfield  {author} {\bibinfo {author} {\bibfnamefont {V.}~\bibnamefont
      {Seshadri}}\ and\ \bibinfo {author} {\bibfnamefont {K.}~\bibnamefont
      {Lindenberg}},\ }\href {https://doi.org/10.1016/0378-4371(82)90036-X}
      {\bibfield  {journal} {\bibinfo  {journal} {Physica A}\ }\textbf {\bibinfo
      {volume} {115}},\ \bibinfo {pages} {501} (\bibinfo {year}
      {1982})}\BibitemShut {NoStop}%
    \bibitem [{\citenamefont
      {Jayannavar}(1991)}]{jayannavarBrownianMotionSpins1991}%
      \BibitemOpen
      \bibfield  {author} {\bibinfo {author} {\bibfnamefont {A.~M.}\ \bibnamefont
      {Jayannavar}},\ }\href {https://doi.org/10.1007/BF01313998} {\bibfield
      {journal} {\bibinfo  {journal} {Z. Physik B - Condensed Matter}\ }\textbf
      {\bibinfo {volume} {82}},\ \bibinfo {pages} {153} (\bibinfo {year}
      {1991})}\BibitemShut {NoStop}%
    \bibitem [{\citenamefont {Rossi}\ \emph {et~al.}(2005)\citenamefont {Rossi},
      \citenamefont {Heinonen},\ and\ \citenamefont
      {MacDonald}}]{rossiDynamicsMagnetizationCoupled2005}%
      \BibitemOpen
      \bibfield  {author} {\bibinfo {author} {\bibfnamefont {E.}~\bibnamefont
      {Rossi}}, \bibinfo {author} {\bibfnamefont {O.~G.}\ \bibnamefont
      {Heinonen}},\ and\ \bibinfo {author} {\bibfnamefont {A.~H.}\ \bibnamefont
      {MacDonald}},\ }\href {https://doi.org/10.1103/PhysRevB.72.174412} {\bibfield
       {journal} {\bibinfo  {journal} {Phys. Rev. B}\ }\textbf {\bibinfo {volume}
      {72}},\ \bibinfo {pages} {174412} (\bibinfo {year} {2005})}\BibitemShut
      {NoStop}%
    \bibitem [{\citenamefont {Vittoria}\ \emph {et~al.}(2010)\citenamefont
      {Vittoria}, \citenamefont {Yoon},\ and\ \citenamefont
      {Widom}}]{vittoriaRelaxationMechanismOrdered2010}%
      \BibitemOpen
      \bibfield  {author} {\bibinfo {author} {\bibfnamefont {C.}~\bibnamefont
      {Vittoria}}, \bibinfo {author} {\bibfnamefont {S.~D.}\ \bibnamefont {Yoon}},\
      and\ \bibinfo {author} {\bibfnamefont {A.}~\bibnamefont {Widom}},\ }\href
      {https://doi.org/10.1103/PhysRevB.81.014412} {\bibfield  {journal} {\bibinfo
      {journal} {Phys. Rev. B}\ }\textbf {\bibinfo {volume} {81}},\ \bibinfo
      {pages} {014412} (\bibinfo {year} {2010})}\BibitemShut {NoStop}%
    \bibitem [{\citenamefont {Miyazaki}\ and\ \citenamefont
      {Seki}(1998)}]{miyazakiBrownianMotionSpins1998}%
      \BibitemOpen
      \bibfield  {author} {\bibinfo {author} {\bibfnamefont {K.}~\bibnamefont
      {Miyazaki}}\ and\ \bibinfo {author} {\bibfnamefont {K.}~\bibnamefont
      {Seki}},\ }\href {https://doi.org/10.1063/1.476123} {\bibfield  {journal}
      {\bibinfo  {journal} {J. Chem. Phys.}\ }\textbf {\bibinfo {volume} {108}},\
      \bibinfo {pages} {7052} (\bibinfo {year} {1998})}\BibitemShut {NoStop}%
    \bibitem [{\citenamefont {Bose}\ and\ \citenamefont
      {Trimper}(2011)}]{boseRetardationEffectsLandauLifshitzGilbert2011}%
      \BibitemOpen
      \bibfield  {author} {\bibinfo {author} {\bibfnamefont {T.}~\bibnamefont
      {Bose}}\ and\ \bibinfo {author} {\bibfnamefont {S.}~\bibnamefont {Trimper}},\
      }\href {https://doi.org/10.1103/PhysRevB.83.134434} {\bibfield  {journal}
      {\bibinfo  {journal} {Phys. Rev. B}\ }\textbf {\bibinfo {volume} {83}},\
      \bibinfo {pages} {134434} (\bibinfo {year} {2011})}\BibitemShut {NoStop}%
    \bibitem [{\citenamefont {Anders}\ \emph {et~al.}(2021)\citenamefont {Anders},
      \citenamefont {Sait},\ and\ \citenamefont
      {Horsley}}]{andersQuantumBrownianMotion2021}%
      \BibitemOpen
      \bibfield  {author} {\bibinfo {author} {\bibfnamefont {J.}~\bibnamefont
      {Anders}}, \bibinfo {author} {\bibfnamefont {C.~R.~J.}\ \bibnamefont
      {Sait}},\ and\ \bibinfo {author} {\bibfnamefont {S.~A.~R.}\ \bibnamefont
      {Horsley}},\ }\href@noop {} {\bibfield  {journal} {\bibinfo  {journal}
      {arXiv:2009.00600 [cond-mat, physics:quant-ph]}\ } (\bibinfo {year}
      {2021})},\ \Eprint {https://arxiv.org/abs/2009.00600} {arXiv:2009.00600
      [cond-mat, physics:quant-ph]} \BibitemShut {NoStop}%
    \bibitem [{\citenamefont {Nieves}\ \emph {et~al.}(2014)\citenamefont {Nieves},
      \citenamefont {Serantes}, \citenamefont {Atxitia},\ and\ \citenamefont
      {{Chubykalo-Fesenko}}}]{nievesQuantumLandauLifshitzBlochEquation2014}%
      \BibitemOpen
      \bibfield  {author} {\bibinfo {author} {\bibfnamefont {P.}~\bibnamefont
      {Nieves}}, \bibinfo {author} {\bibfnamefont {D.}~\bibnamefont {Serantes}},
      \bibinfo {author} {\bibfnamefont {U.}~\bibnamefont {Atxitia}},\ and\ \bibinfo
      {author} {\bibfnamefont {O.}~\bibnamefont {{Chubykalo-Fesenko}}},\ }\href
      {https://doi.org/10.1103/PhysRevB.90.104428} {\bibfield  {journal} {\bibinfo
      {journal} {Phys. Rev. B}\ }\textbf {\bibinfo {volume} {90}},\ \bibinfo
      {pages} {104428} (\bibinfo {year} {2014})}\BibitemShut {NoStop}%
    \bibitem [{\citenamefont {Maniv}\ \emph {et~al.}(2021)\citenamefont {Maniv},
      \citenamefont {Nair}, \citenamefont {Haley}, \citenamefont {Doyle},
      \citenamefont {John}, \citenamefont {Cabrini}, \citenamefont {Maniv},
      \citenamefont {Ramakrishna}, \citenamefont {Tang}, \citenamefont {Ercius},
      \citenamefont {Ramesh}, \citenamefont {Tserkovnyak}, \citenamefont {Reyes},\
      and\ \citenamefont {Analytis}}]{manivAntiferromagneticSwitchingDriven2021}%
      \BibitemOpen
      \bibfield  {author} {\bibinfo {author} {\bibfnamefont {E.}~\bibnamefont
      {Maniv}}, \bibinfo {author} {\bibfnamefont {N.~L.}\ \bibnamefont {Nair}},
      \bibinfo {author} {\bibfnamefont {S.~C.}\ \bibnamefont {Haley}}, \bibinfo
      {author} {\bibfnamefont {S.}~\bibnamefont {Doyle}}, \bibinfo {author}
      {\bibfnamefont {C.}~\bibnamefont {John}}, \bibinfo {author} {\bibfnamefont
      {S.}~\bibnamefont {Cabrini}}, \bibinfo {author} {\bibfnamefont
      {A.}~\bibnamefont {Maniv}}, \bibinfo {author} {\bibfnamefont {S.~K.}\
      \bibnamefont {Ramakrishna}}, \bibinfo {author} {\bibfnamefont {Y.-L.}\
      \bibnamefont {Tang}}, \bibinfo {author} {\bibfnamefont {P.}~\bibnamefont
      {Ercius}}, \bibinfo {author} {\bibfnamefont {R.}~\bibnamefont {Ramesh}},
      \bibinfo {author} {\bibfnamefont {Y.}~\bibnamefont {Tserkovnyak}}, \bibinfo
      {author} {\bibfnamefont {A.~P.}\ \bibnamefont {Reyes}},\ and\ \bibinfo
      {author} {\bibfnamefont {J.~G.}\ \bibnamefont {Analytis}},\ }\href
      {https://doi.org/10.1126/sciadv.abd8452} {\bibfield  {journal} {\bibinfo
      {journal} {Science Advances}\ }\textbf {\bibinfo {volume} {7}},\ \bibinfo
      {pages} {eabd8452} (\bibinfo {year} {2021})}\BibitemShut {NoStop}%
    \bibitem [{\citenamefont {Belletti}\ \emph {et~al.}(2008)\citenamefont
      {Belletti}, \citenamefont {Cotallo}, \citenamefont {Cruz}, \citenamefont
      {Fernandez}, \citenamefont {{Gordillo-Guerrero}}, \citenamefont {Guidetti},
      \citenamefont {Maiorano}, \citenamefont {Mantovani}, \citenamefont
      {Marinari}, \citenamefont {{Martin-Mayor}}, \citenamefont {Sudupe},
      \citenamefont {Navarro}, \citenamefont {Parisi}, \citenamefont
      {{Perez-Gaviro}}, \citenamefont {{Ruiz-Lorenzo}}, \citenamefont {Schifano},
      \citenamefont {Sciretti}, \citenamefont {Tarancon}, \citenamefont
      {Tripiccione}, \citenamefont {Velasco},\ and\ \citenamefont
      {Yllanes}}]{bellettiNonequilibriumSpinGlassDynamics2008}%
      \BibitemOpen
      \bibfield  {author} {\bibinfo {author} {\bibfnamefont {F.}~\bibnamefont
      {Belletti}}, \bibinfo {author} {\bibfnamefont {M.}~\bibnamefont {Cotallo}},
      \bibinfo {author} {\bibfnamefont {A.}~\bibnamefont {Cruz}}, \bibinfo {author}
      {\bibfnamefont {L.~A.}\ \bibnamefont {Fernandez}}, \bibinfo {author}
      {\bibfnamefont {A.}~\bibnamefont {{Gordillo-Guerrero}}}, \bibinfo {author}
      {\bibfnamefont {M.}~\bibnamefont {Guidetti}}, \bibinfo {author}
      {\bibfnamefont {A.}~\bibnamefont {Maiorano}}, \bibinfo {author}
      {\bibfnamefont {F.}~\bibnamefont {Mantovani}}, \bibinfo {author}
      {\bibfnamefont {E.}~\bibnamefont {Marinari}}, \bibinfo {author}
      {\bibfnamefont {V.}~\bibnamefont {{Martin-Mayor}}}, \bibinfo {author}
      {\bibfnamefont {A.~M.}\ \bibnamefont {Sudupe}}, \bibinfo {author}
      {\bibfnamefont {D.}~\bibnamefont {Navarro}}, \bibinfo {author} {\bibfnamefont
      {G.}~\bibnamefont {Parisi}}, \bibinfo {author} {\bibfnamefont
      {S.}~\bibnamefont {{Perez-Gaviro}}}, \bibinfo {author} {\bibfnamefont
      {J.~J.}\ \bibnamefont {{Ruiz-Lorenzo}}}, \bibinfo {author} {\bibfnamefont
      {S.~F.}\ \bibnamefont {Schifano}}, \bibinfo {author} {\bibfnamefont
      {D.}~\bibnamefont {Sciretti}}, \bibinfo {author} {\bibfnamefont
      {A.}~\bibnamefont {Tarancon}}, \bibinfo {author} {\bibfnamefont
      {R.}~\bibnamefont {Tripiccione}}, \bibinfo {author} {\bibfnamefont {J.~L.}\
      \bibnamefont {Velasco}},\ and\ \bibinfo {author} {\bibfnamefont
      {D.}~\bibnamefont {Yllanes}},\ }\href
      {https://doi.org/10.1103/PhysRevLett.101.157201} {\bibfield  {journal}
      {\bibinfo  {journal} {Phys. Rev. Lett.}\ }\textbf {\bibinfo {volume} {101}},\
      \bibinfo {pages} {157201} (\bibinfo {year} {2008})}\BibitemShut {NoStop}%
    \bibitem [{\citenamefont {Ashworth}\ and\ \citenamefont
      {Davies}(1979)}]{ashworthTransformationsInertialRotating1979}%
      \BibitemOpen
      \bibfield  {author} {\bibinfo {author} {\bibfnamefont {D.~G.}\ \bibnamefont
      {Ashworth}}\ and\ \bibinfo {author} {\bibfnamefont {P.~A.}\ \bibnamefont
      {Davies}},\ }\href {https://doi.org/10.1088/0305-4470/12/9/011} {\bibfield
      {journal} {\bibinfo  {journal} {J. Phys. A: Math. Gen.}\ }\textbf {\bibinfo
      {volume} {12}},\ \bibinfo {pages} {1425} (\bibinfo {year}
      {1979})}\BibitemShut {NoStop}%
    \bibitem [{\citenamefont {Rubi}\ and\ \citenamefont
      {{Perez-Madrid}}(1999)}]{rubiInertialEffectsNonequilibrium1999}%
      \BibitemOpen
      \bibfield  {author} {\bibinfo {author} {\bibfnamefont {J.~M.}\ \bibnamefont
      {Rubi}}\ and\ \bibinfo {author} {\bibfnamefont {A.}~\bibnamefont
      {{Perez-Madrid}}},\ }\href {https://doi.org/10.1016/S0378-4371(98)00476-2}
      {\bibfield  {journal} {\bibinfo  {journal} {Physica A}\ }\textbf {\bibinfo
      {volume} {264}},\ \bibinfo {pages} {492} (\bibinfo {year}
      {1999})}\BibitemShut {NoStop}%
    \bibitem [{\citenamefont {Ciornei}\ \emph {et~al.}(2011)\citenamefont
      {Ciornei}, \citenamefont {Rub{\'i}},\ and\ \citenamefont
      {Wegrowe}}]{ciorneiMagnetizationDynamicsInertial2011}%
      \BibitemOpen
      \bibfield  {author} {\bibinfo {author} {\bibfnamefont {M.-C.}\ \bibnamefont
      {Ciornei}}, \bibinfo {author} {\bibfnamefont {J.~M.}\ \bibnamefont
      {Rub{\'i}}},\ and\ \bibinfo {author} {\bibfnamefont {J.-E.}\ \bibnamefont
      {Wegrowe}},\ }\href {https://doi.org/10.1103/PhysRevB.83.020410} {\bibfield
      {journal} {\bibinfo  {journal} {Phys. Rev. B}\ }\textbf {\bibinfo {volume}
      {83}},\ \bibinfo {pages} {020410(R)} (\bibinfo {year} {2011})}\BibitemShut
      {NoStop}%
    \bibitem [{\citenamefont {Bhattacharjee}\ \emph {et~al.}(2012)\citenamefont
      {Bhattacharjee}, \citenamefont {Nordstr{\"o}m},\ and\ \citenamefont
      {Fransson}}]{bhattacharjeeAtomisticSpinDynamic2012}%
      \BibitemOpen
      \bibfield  {author} {\bibinfo {author} {\bibfnamefont {S.}~\bibnamefont
      {Bhattacharjee}}, \bibinfo {author} {\bibfnamefont {L.}~\bibnamefont
      {Nordstr{\"o}m}},\ and\ \bibinfo {author} {\bibfnamefont {J.}~\bibnamefont
      {Fransson}},\ }\href {https://doi.org/10.1103/PhysRevLett.108.057204}
      {\bibfield  {journal} {\bibinfo  {journal} {Phys. Rev. Lett.}\ }\textbf
      {\bibinfo {volume} {108}},\ \bibinfo {pages} {057204} (\bibinfo {year}
      {2012})}\BibitemShut {NoStop}%
    \bibitem [{\citenamefont {Pervishko}\ \emph {et~al.}(2018)\citenamefont
      {Pervishko}, \citenamefont {Baglai}, \citenamefont {Eriksson},\ and\
      \citenamefont {Yudin}}]{pervishkoAnotherViewGilbert2018}%
      \BibitemOpen
      \bibfield  {author} {\bibinfo {author} {\bibfnamefont {A.~A.}\ \bibnamefont
      {Pervishko}}, \bibinfo {author} {\bibfnamefont {M.~I.}\ \bibnamefont
      {Baglai}}, \bibinfo {author} {\bibfnamefont {O.}~\bibnamefont {Eriksson}},\
      and\ \bibinfo {author} {\bibfnamefont {D.}~\bibnamefont {Yudin}},\ }\href
      {https://doi.org/10.1038/s41598-018-35517-x} {\bibfield  {journal} {\bibinfo
      {journal} {Sci. Rep.}\ }\textbf {\bibinfo {volume} {8}},\ \bibinfo {pages}
      {17148} (\bibinfo {year} {2018})}\BibitemShut {NoStop}%
    \bibitem [{\citenamefont {Mondal}\ \emph {et~al.}(2018)\citenamefont {Mondal},
      \citenamefont {Berritta},\ and\ \citenamefont
      {Oppeneer}}]{mondalGeneralisationGilbertDamping2018}%
      \BibitemOpen
      \bibfield  {author} {\bibinfo {author} {\bibfnamefont {R.}~\bibnamefont
      {Mondal}}, \bibinfo {author} {\bibfnamefont {M.}~\bibnamefont {Berritta}},\
      and\ \bibinfo {author} {\bibfnamefont {P.~M.}\ \bibnamefont {Oppeneer}},\
      }\href {https://doi.org/10.1088/1361-648X/aac5a2} {\bibfield  {journal}
      {\bibinfo  {journal} {J. Phys.: Condens. Matter}\ }\textbf {\bibinfo {volume}
      {30}},\ \bibinfo {pages} {265801} (\bibinfo {year} {2018})}\BibitemShut
      {NoStop}%
    \bibitem [{\citenamefont {N{\'e}el}(1971)}]{neelMagnetismLocalMolecular1971}%
      \BibitemOpen
      \bibfield  {author} {\bibinfo {author} {\bibfnamefont {L.}~\bibnamefont
      {N{\'e}el}},\ }\href {https://doi.org/10.1126/science.174.4013.985}
      {\bibfield  {journal} {\bibinfo  {journal} {Science}\ }\textbf {\bibinfo
      {volume} {174}},\ \bibinfo {pages} {985} (\bibinfo {year}
      {1971})}\BibitemShut {NoStop}%
    \bibitem [{\citenamefont {Guimar{\~a}es}\ \emph {et~al.}(2019)\citenamefont
      {Guimar{\~a}es}, \citenamefont {Suckert}, \citenamefont {Chico},
      \citenamefont {Bouaziz}, \citenamefont {Dias},\ and\ \citenamefont
      {Lounis}}]{guimaraesComparativeStudyMethodologies2019}%
      \BibitemOpen
      \bibfield  {author} {\bibinfo {author} {\bibfnamefont {F.~S.~M.}\
      \bibnamefont {Guimar{\~a}es}}, \bibinfo {author} {\bibfnamefont {J.~R.}\
      \bibnamefont {Suckert}}, \bibinfo {author} {\bibfnamefont {J.}~\bibnamefont
      {Chico}}, \bibinfo {author} {\bibfnamefont {J.}~\bibnamefont {Bouaziz}},
      \bibinfo {author} {\bibfnamefont {M.~d.~S.}\ \bibnamefont {Dias}},\ and\
      \bibinfo {author} {\bibfnamefont {S.}~\bibnamefont {Lounis}},\ }\href
      {https://doi.org/10.1088/1361-648X/ab1239} {\bibfield  {journal} {\bibinfo
      {journal} {J. Phys.: Condens. Matter}\ }\textbf {\bibinfo {volume} {31}},\
      \bibinfo {pages} {255802} (\bibinfo {year} {2019})}\BibitemShut {NoStop}%
    \bibitem [{\citenamefont
      {Kambersk{\'y}}(1970)}]{kamberskyLandauLifshitzRelaxation1970}%
      \BibitemOpen
      \bibfield  {author} {\bibinfo {author} {\bibfnamefont {V.}~\bibnamefont
      {Kambersk{\'y}}},\ }\href {https://doi.org/10.1139/p70-361} {\bibfield
      {journal} {\bibinfo  {journal} {Can. J. Phys.}\ }\textbf {\bibinfo {volume}
      {48}},\ \bibinfo {pages} {2906} (\bibinfo {year} {1970})}\BibitemShut
      {NoStop}%
    \bibitem [{\citenamefont
      {Zwanzig}(2001)}]{zwanzigNonequilibriumStatisticalMechanics2001}%
      \BibitemOpen
      \bibfield  {author} {\bibinfo {author} {\bibfnamefont {R.}~\bibnamefont
      {Zwanzig}},\ }\href@noop {} {{\selectlanguage {English}\emph {\bibinfo
      {title} {Nonequilibrium Statistical Mechanics}}}}\ (\bibinfo  {publisher}
      {{Oxford Univ. Press}},\ \bibinfo {address} {{Oxford}},\ \bibinfo {year}
      {2001})\BibitemShut {NoStop}%
    \bibitem [{\citenamefont {Parisi}\ and\ \citenamefont
      {Sourlas}(1982)}]{parisiSupersymmetricFieldTheories1982}%
      \BibitemOpen
      \bibfield  {author} {\bibinfo {author} {\bibfnamefont {G.}~\bibnamefont
      {Parisi}}\ and\ \bibinfo {author} {\bibfnamefont {N.}~\bibnamefont
      {Sourlas}},\ }\href {https://doi.org/10.1016/0550-3213(82)90538-7} {\bibfield
       {journal} {\bibinfo  {journal} {Nucl. Phys. B}\ }\textbf {\bibinfo {volume}
      {206}},\ \bibinfo {pages} {321} (\bibinfo {year} {1982})}\BibitemShut
      {NoStop}%
    \bibitem [{\citenamefont {Nicolis}(2019)}]{nicolisParticleEquilibriumBath2019}%
      \BibitemOpen
      \bibfield  {author} {\bibinfo {author} {\bibfnamefont {S.}~\bibnamefont
      {Nicolis}},\ }\href@noop {} {\bibfield  {journal} {\bibinfo  {journal}
      {arXiv:1405.0820}\ } (\bibinfo {year} {2019})},\ \Eprint
      {https://arxiv.org/abs/1405.0820} {arXiv:1405.0820} \BibitemShut {NoStop}%
    \bibitem [{\citenamefont {Atxitia}\ \emph {et~al.}(2009)\citenamefont
      {Atxitia}, \citenamefont {{Chubykalo-Fesenko}}, \citenamefont {Chantrell},
      \citenamefont {Nowak},\ and\ \citenamefont
      {Rebei}}]{atxitiaUltrafastSpinDynamics2009}%
      \BibitemOpen
      \bibfield  {author} {\bibinfo {author} {\bibfnamefont {U.}~\bibnamefont
      {Atxitia}}, \bibinfo {author} {\bibfnamefont {O.}~\bibnamefont
      {{Chubykalo-Fesenko}}}, \bibinfo {author} {\bibfnamefont {R.~W.}\
      \bibnamefont {Chantrell}}, \bibinfo {author} {\bibfnamefont {U.}~\bibnamefont
      {Nowak}},\ and\ \bibinfo {author} {\bibfnamefont {A.}~\bibnamefont {Rebei}},\
      }\href {https://doi.org/10.1103/PhysRevLett.102.057203} {\bibfield  {journal}
      {\bibinfo  {journal} {Phys. Rev. Lett.}\ }\textbf {\bibinfo {volume} {102}},\
      \bibinfo {pages} {057203} (\bibinfo {year} {2009})}\BibitemShut {NoStop}%
    \bibitem [{\citenamefont
      {{Zinn-Justin}}(2007)}]{zinn-justinPhaseTransitionsRenormalization2007}%
      \BibitemOpen
      \bibfield  {author} {\bibinfo {author} {\bibfnamefont {J.}~\bibnamefont
      {{Zinn-Justin}}},\ }\href@noop {} {{\selectlanguage {English}\emph {\bibinfo
      {title} {Phase {{Transitions}} and {{Renormalization Group}}}}}}\ (\bibinfo
      {publisher} {{Oxford University Press}},\ \bibinfo {address} {{Oxford}},\
      \bibinfo {year} {2007})\BibitemShut {NoStop}%
    \bibitem [{\citenamefont
      {{Zinn-Justin}}(2002)}]{zinn-justinQuantumFieldTheory2002}%
      \BibitemOpen
      \bibfield  {author} {\bibinfo {author} {\bibfnamefont {J.}~\bibnamefont
      {{Zinn-Justin}}},\ }\href@noop {} {{\selectlanguage {English}\emph {\bibinfo
      {title} {Quantum Field Theory and Critical Phenomena}}}},\ \bibinfo {edition}
      {4th}\ ed.,\ \bibinfo {series} {International Series of Monographs on
      Physics}\ No.\ \bibinfo {number} {113}\ (\bibinfo  {publisher} {{Clarendon
      Press ; Oxford University Press}},\ \bibinfo {address} {{Oxford : New
      York}},\ \bibinfo {year} {2002})\BibitemShut {NoStop}%
    \bibitem [{\citenamefont {Cugliandolo}\ and\ \citenamefont
      {Lecomte}(2017)}]{cugliandoloRulesCalculusPath2017}%
      \BibitemOpen
      \bibfield  {author} {\bibinfo {author} {\bibfnamefont {L.~F.}\ \bibnamefont
      {Cugliandolo}}\ and\ \bibinfo {author} {\bibfnamefont {V.}~\bibnamefont
      {Lecomte}},\ }\href {https://doi.org/10.1088/1751-8121/aa7dd6} {\bibfield
      {journal} {\bibinfo  {journal} {J. Phys. A: Math. Theor.}\ }\textbf {\bibinfo
      {volume} {50}},\ \bibinfo {pages} {345001} (\bibinfo {year}
      {2017})}\BibitemShut {NoStop}%
    \bibitem [{\citenamefont
      {{Zinn-Justin}}(2009)}]{zinn-justinPathIntegralsQuantum2009}%
      \BibitemOpen
      \bibfield  {author} {\bibinfo {author} {\bibfnamefont {J.}~\bibnamefont
      {{Zinn-Justin}}},\ }\href@noop {} {{\selectlanguage {English}\emph {\bibinfo
      {title} {Path Integrals in Quantum Mechanics}}}},\ \bibinfo {edition}
      {reprint}\ ed.,\ Oxford Graduate Texts\ (\bibinfo  {publisher} {{Oxford Univ.
      Press}},\ \bibinfo {address} {{Oxford}},\ \bibinfo {year} {2009})\BibitemShut
      {NoStop}%
    \bibitem [{\citenamefont
      {Gardiner}(2009)}]{gardinerStochasticMethodsHandbook2009}%
      \BibitemOpen
      \bibfield  {author} {\bibinfo {author} {\bibfnamefont {C.~W.}\ \bibnamefont
      {Gardiner}},\ }\href@noop {} {{\selectlanguage {English}\emph {\bibinfo
      {title} {Stochastic Methods: A Handbook for the Natural and Social
      Sciences}}}},\ \bibinfo {edition} {4th}\ ed.,\ Springer Series in
      Synergetics\ (\bibinfo  {publisher} {{Springer}},\ \bibinfo {address}
      {{Berlin}},\ \bibinfo {year} {2009})\BibitemShut {NoStop}%
    \bibitem [{\citenamefont
      {Van~Kampen}(1992)}]{vankampenStochasticProcessesPhysics1992}%
      \BibitemOpen
      \bibfield  {author} {\bibinfo {author} {\bibfnamefont {N.~G.~V.}\
      \bibnamefont {Van~Kampen}},\ }\href@noop {} {{\selectlanguage {English}\emph
      {\bibinfo {title} {Stochastic {{Processes}} in {{Physics}} and
      {{Chemistry}}}}}}\ (\bibinfo  {publisher} {{Elsevier}},\ \bibinfo {year}
      {1992})\BibitemShut {NoStop}%
    \bibitem [{\citenamefont {Tranchida}\ \emph {et~al.}(2018)\citenamefont
      {Tranchida}, \citenamefont {Thibaudeau},\ and\ \citenamefont
      {Nicolis}}]{tranchidaHierarchiesLandauLifshitzBlochEquations2018}%
      \BibitemOpen
      \bibfield  {author} {\bibinfo {author} {\bibfnamefont {J.}~\bibnamefont
      {Tranchida}}, \bibinfo {author} {\bibfnamefont {P.}~\bibnamefont
      {Thibaudeau}},\ and\ \bibinfo {author} {\bibfnamefont {S.}~\bibnamefont
      {Nicolis}},\ }\href {https://doi.org/10.1103/PhysRevE.98.042101} {\bibfield
      {journal} {\bibinfo  {journal} {Phys. Rev. E}\ }\textbf {\bibinfo {volume}
      {98}},\ \bibinfo {pages} {042101} (\bibinfo {year} {2018})}\BibitemShut
      {NoStop}%
    \bibitem [{\citenamefont {Weir}(1973)}]{weirLebesgueIntegrationMeasure1973}%
      \BibitemOpen
      \bibfield  {author} {\bibinfo {author} {\bibfnamefont {A.~J.}\ \bibnamefont
      {Weir}},\ }\href@noop {} {{\selectlanguage {English}\emph {\bibinfo {title}
      {Lebesgue Integration and Measure}}}}\ (\bibinfo  {publisher} {{University
      Press}},\ \bibinfo {address} {{Cambridge [Eng.]}},\ \bibinfo {year}
      {1973})\BibitemShut {NoStop}%
    \bibitem [{\citenamefont {Garate}\ and\ \citenamefont
      {MacDonald}(2009)}]{garateGilbertDampingConducting2009}%
      \BibitemOpen
      \bibfield  {author} {\bibinfo {author} {\bibfnamefont {I.}~\bibnamefont
      {Garate}}\ and\ \bibinfo {author} {\bibfnamefont {A.}~\bibnamefont
      {MacDonald}},\ }\href {https://doi.org/10.1103/PhysRevB.79.064403} {\bibfield
       {journal} {\bibinfo  {journal} {Phys. Rev. B}\ }\textbf {\bibinfo {volume}
      {79}},\ \bibinfo {pages} {064403} (\bibinfo {year} {2009})}\BibitemShut
      {NoStop}%
    \bibitem [{\citenamefont {Garate}\ \emph {et~al.}(2009)\citenamefont {Garate},
      \citenamefont {Gilmore}, \citenamefont {Stiles},\ and\ \citenamefont
      {MacDonald}}]{garateNonadiabaticSpintransferTorque2009}%
      \BibitemOpen
      \bibfield  {author} {\bibinfo {author} {\bibfnamefont {I.}~\bibnamefont
      {Garate}}, \bibinfo {author} {\bibfnamefont {K.}~\bibnamefont {Gilmore}},
      \bibinfo {author} {\bibfnamefont {M.~D.}\ \bibnamefont {Stiles}},\ and\
      \bibinfo {author} {\bibfnamefont {A.~H.}\ \bibnamefont {MacDonald}},\ }\href
      {https://doi.org/10.1103/PhysRevB.79.104416} {\bibfield  {journal} {\bibinfo
      {journal} {Phys. Rev. B}\ }\textbf {\bibinfo {volume} {79}},\ \bibinfo
      {pages} {104416} (\bibinfo {year} {2009})}\BibitemShut {NoStop}%
    \bibitem [{\citenamefont {Thonig}\ \emph {et~al.}(2018)\citenamefont {Thonig},
      \citenamefont {Kvashnin}, \citenamefont {Eriksson},\ and\ \citenamefont
      {Pereiro}}]{thonigNonlocalGilbertDamping2018}%
      \BibitemOpen
      \bibfield  {author} {\bibinfo {author} {\bibfnamefont {D.}~\bibnamefont
      {Thonig}}, \bibinfo {author} {\bibfnamefont {Y.}~\bibnamefont {Kvashnin}},
      \bibinfo {author} {\bibfnamefont {O.}~\bibnamefont {Eriksson}},\ and\
      \bibinfo {author} {\bibfnamefont {M.}~\bibnamefont {Pereiro}},\ }\href
      {https://doi.org/10.1103/PhysRevMaterials.2.013801} {\bibfield  {journal}
      {\bibinfo  {journal} {Phys. Rev. Materials}\ }\textbf {\bibinfo {volume}
      {2}},\ \bibinfo {pages} {013801} (\bibinfo {year} {2018})}\BibitemShut
      {NoStop}%
    \bibitem [{\citenamefont {Hickey}\ and\ \citenamefont
      {Moodera}(2009)}]{hickeyOriginIntrinsicGilbert2009}%
      \BibitemOpen
      \bibfield  {author} {\bibinfo {author} {\bibfnamefont {M.~C.}\ \bibnamefont
      {Hickey}}\ and\ \bibinfo {author} {\bibfnamefont {J.~S.}\ \bibnamefont
      {Moodera}},\ }\href {https://doi.org/10.1103/PhysRevLett.102.137601}
      {\bibfield  {journal} {\bibinfo  {journal} {Phys. Rev. Lett.}\ }\textbf
      {\bibinfo {volume} {102}},\ \bibinfo {pages} {137601} (\bibinfo {year}
      {2009})}\BibitemShut {NoStop}%
    \bibitem [{\citenamefont {F{\"a}hnle}\ \emph {et~al.}(2011)\citenamefont
      {F{\"a}hnle}, \citenamefont {Steiauf},\ and\ \citenamefont
      {Illg}}]{fahnleGeneralizedGilbertEquation2011}%
      \BibitemOpen
      \bibfield  {author} {\bibinfo {author} {\bibfnamefont {M.}~\bibnamefont
      {F{\"a}hnle}}, \bibinfo {author} {\bibfnamefont {D.}~\bibnamefont
      {Steiauf}},\ and\ \bibinfo {author} {\bibfnamefont {C.}~\bibnamefont
      {Illg}},\ }\href {https://doi.org/10.1103/PhysRevB.84.172403} {\bibfield
      {journal} {\bibinfo  {journal} {Phys. Rev. B}\ }\textbf {\bibinfo {volume}
      {84}},\ \bibinfo {pages} {172403} (\bibinfo {year} {2011})}\BibitemShut
      {NoStop}%
    \bibitem [{\citenamefont {F{\"a}hnle}\ \emph {et~al.}(2013)\citenamefont
      {F{\"a}hnle}, \citenamefont {Steiauf},\ and\ \citenamefont
      {Illg}}]{fahnleErratumGeneralizedGilbert2013}%
      \BibitemOpen
      \bibfield  {author} {\bibinfo {author} {\bibfnamefont {M.}~\bibnamefont
      {F{\"a}hnle}}, \bibinfo {author} {\bibfnamefont {D.}~\bibnamefont
      {Steiauf}},\ and\ \bibinfo {author} {\bibfnamefont {C.}~\bibnamefont
      {Illg}},\ }\bibfield  {journal} {\bibinfo  {journal} {Phys. Rev. B}\ }\textbf
      {\bibinfo {volume} {88}},\ \href {https://doi.org/10.1103/PhysRevB.88.219905}
      {10.1103/PhysRevB.88.219905} (\bibinfo {year} {2013})\BibitemShut {NoStop}%
    \bibitem [{\citenamefont {Suhl}(1998)}]{suhlTheoryMagneticDamping1998}%
      \BibitemOpen
      \bibfield  {author} {\bibinfo {author} {\bibfnamefont {H.}~\bibnamefont
      {Suhl}},\ }\href {https://doi.org/10.1109/20.706720} {\bibfield  {journal}
      {\bibinfo  {journal} {IEEE Trans. Magn.}\ }\textbf {\bibinfo {volume} {34}},\
      \bibinfo {pages} {1834} (\bibinfo {year} {1998})}\BibitemShut {NoStop}%
    \bibitem [{\citenamefont
      {Suhl}(2007)}]{suhlRelaxationProcessesMicromagnetics2007}%
      \BibitemOpen
      \bibfield  {author} {\bibinfo {author} {\bibfnamefont {H.}~\bibnamefont
      {Suhl}},\ }\href {https://doi.org/10.1093/acprof:oso/9780198528029.001.0001}
      {{\selectlanguage {English}\emph {\bibinfo {title} {Relaxation {{Processes}}
      in {{Micromagnetics}}}}}}\ (\bibinfo  {publisher} {{Oxford University
      Press}},\ \bibinfo {address} {{Oxford}},\ \bibinfo {year} {2007})\BibitemShut
      {NoStop}%
    \bibitem [{\citenamefont {Omelyan}\ \emph {et~al.}(2003)\citenamefont
      {Omelyan}, \citenamefont {Mryglod},\ and\ \citenamefont
      {Folk}}]{omelyanSymplecticAnalyticallyIntegrable2003}%
      \BibitemOpen
      \bibfield  {author} {\bibinfo {author} {\bibfnamefont {I.~P.}\ \bibnamefont
      {Omelyan}}, \bibinfo {author} {\bibfnamefont {I.~M.}\ \bibnamefont
      {Mryglod}},\ and\ \bibinfo {author} {\bibfnamefont {R.}~\bibnamefont
      {Folk}},\ }\href {https://doi.org/10.1016/S0010-4655(02)00754-3} {\bibfield
      {journal} {\bibinfo  {journal} {Comp. Phys. Comm.}\ }\textbf {\bibinfo
      {volume} {151}},\ \bibinfo {pages} {272} (\bibinfo {year}
      {2003})}\BibitemShut {NoStop}%
    \bibitem [{\citenamefont {Nicolis}\ \emph {et~al.}(2017)\citenamefont
      {Nicolis}, \citenamefont {Thibaudeau},\ and\ \citenamefont
      {Tranchida}}]{nicolisFinitedimensionalColoredFluctuationdissipation2017}%
      \BibitemOpen
      \bibfield  {author} {\bibinfo {author} {\bibfnamefont {S.}~\bibnamefont
      {Nicolis}}, \bibinfo {author} {\bibfnamefont {P.}~\bibnamefont
      {Thibaudeau}},\ and\ \bibinfo {author} {\bibfnamefont {J.}~\bibnamefont
      {Tranchida}},\ }\href {https://doi.org/10.1063/1.4975132} {\bibfield
      {journal} {\bibinfo  {journal} {AIP Adv.}\ }\textbf {\bibinfo {volume} {7}},\
      \bibinfo {pages} {056012} (\bibinfo {year} {2017})}\BibitemShut {NoStop}%
    \end{thebibliography}
%
\end{document}